\documentclass[useAMS,usenatbib,referee]{biom}

\usepackage{subcaption}

\usepackage{booktabs,tabularx}
\usepackage{float}
\usepackage{xcolor}
\usepackage{soul,color}
\usepackage[figuresright]{rotating}
\usepackage{amsmath}
\usepackage{lastpage}
\usepackage{appendix}
\usepackage{comment}
\usepackage{diagbox}
\usepackage{amsfonts}
\usepackage{amssymb}
\usepackage{tabu}
\usepackage{caption}
\usepackage{mdwlist}
\usepackage{xcolor}
\usepackage{tabularx}
\usepackage{pifont}
\usepackage{mathtools}
\usepackage{booktabs}
\usepackage{threeparttable}
\usepackage{chngcntr}
\usepackage{setspace}
\usepackage{rotating}
\usepackage{graphicx}
\usepackage{titlesec}
\usepackage{multirow}
\usepackage{tablefootnote}

\titlespacing*{\section}{0pt}{0.5\baselineskip}{0.2\baselineskip}
\titlespacing*{\subsection}{0pt}{0.2\baselineskip}{0.2\baselineskip}

\newcommand{\argmin}{\mathop{\text{\normalfont{argmin}}}}

\newcommand{\sign}{{\text{sign}}}

\newcommand{\PR}{\mathbb{P}}

\newcommand{\Unif}{\text{Unif}}

\newcommand{\I}{\text{I}}
\allowdisplaybreaks

\title{{On $p$-value combination of independent and frequent signals: asymptotic efficiency and Fisher ensemble
}}
\author
{Yusi Fang\emailx{yuf31@pitt.edu} \\
Department of Biostatistics, University of Pittsburgh,
Pittsburgh, U.S.A. \and
Chung Chang \emailx{cchang@math.nsysu.edu.tw}\\
Department of Applied Mathematics,
National Sun Yat-sen University,
Kaohsiung, Taiwan
\and George Tseng\emailx{ctseng@pitt.edu} \\
Department of Biostatistics, University of Pittsburgh,
Pittsburgh, U.S.A.}

\begin{document}





\label{firstpage}


\begin{abstract}
Combining $p$-values to integrate multiple effects is of long-standing interest in social science and biomedical research. In this paper, we focus on revisiting a classical scenario closely related to meta-analysis, which combines a relatively small (finite and fixed) number of $p$-values while the sample size for generating each $p$-value is large (asymptotically goes to infinity). We evaluate a list of traditional and recently developed modified Fisher's methods to investigate their asymptotic efficiencies and finite-sample numerical performance. The result concludes that Fisher and adaptively weighted Fisher method have top performance and complementary advantages across different proportions of true signals. Consequently, we propose an ensemble method, namely Fisher ensemble, to combine the two top-performing Fisher-related methods using a robust truncated Cauchy ensemble approach.
We show that Fisher ensemble achieves asymptotic Bahadur optimality and integrates the strengths of Fisher and adaptively weighted Fisher methods in simulations. We subsequently extend Fisher ensemble to a variant with emphasized power for concordant effect size directions. A transcriptomic meta-analysis application confirms the theoretical and simulation conclusions, generates intriguing biomarker and pathway findings, and demonstrates the strengths and strategy of using the proposed Fisher ensemble methods.   
\end{abstract}


\maketitle

\section{Introduction}\label{Introduction}
Methods for combining $p$-values are historically of substantial interest in statistics and in applications of many scientific fields to aggregate homogeneous or possibly heterogeneous information from multiple sources. Consider the problem of combining $K$ $p$-values, $\vec{p}=(p_1,\ldots,p_K)$, where $p_i$ is $p$-value of testing $H_0^{(i)}:\theta_i\in \Theta_0^{(i)}$ versus $H_1^{(i)}:\theta_i\in \Theta^{(i)}-\Theta_0^{(i)}$. The global union-intersection test for detecting any signal in the $K$ $p$-values is $H_0:\cap_{1\leq i\leq K} \{ \theta_i\in \Theta_0^{(i)}\}$ versus $H_1: \cup_{1\leq i\leq K} \{\theta_i\in \Theta^{(i)}-\Theta_0^{(i)} \}$. A general strategy is to combine the input $p$-values and form a test statistic for globally testing the existence of any signal. In the literature, three major categories of methods have been developed, depending on the types of input data and signal. The first category considers combination of independent $p$-values, where $K$ is small and fixed (usually 5-30). The sample size $n_i$ ($1\leq i\leq K$) for deriving $p_i$ is large and can asymptotically go to infinity. This first classical scenario is closely related to meta-analysis applications to integrate multiple small effects for increased statistical power. Traditional methods include Fisher's method $T_{\text{\text{Fisher}}}=\sum_{i=1}^K -2\log p_i$\citep{fisher1934statistical} and Stouffer's method $T_{\text{Stouffer}}=\sum_{i=1}^K \Phi^{-1}(1-p_i)$\citep{stouffer1949american} as well as many other transformation selections. The second category considers combining independent, sparse, and weak signals, where a large number of $p$-values are combined ($K\rightarrow\infty$) while only a small number $\ell$ of the $K$ $p$-values ($\ell=K^\beta$ with $0<\beta<\frac{1}{2}$) have weak signals and all remaining $p$-values have no signal. High criticism \citep[denoted by HC test hereafter]{donoho2004higher} and Berk-Jones test \citep[denoted by BJ test hereafter]{berk1979goodness,li2015higher} are two representative methods and are shown to be asymptotically optimal in terms of detection boundary across varying levels of signal sparsity ($0<\beta<\frac{1}{2}$) as $K\rightarrow \infty$. In the third category, integration of $K$ $p$-values with unknown correlation structure and with sparse and weak signals is considered. \cite{liu2020cauchy} and \cite{wilson2019harmonic} proposed Cauchy test (CA) and harmonic mean test (HM), respectively. These methods provide robustness under unknown dependency structure when inference is established under independence assumption and they attain optimal detection boundary for detecting highly sparse signals (with $s=K^\beta$, $0<\beta<\frac{1}{4}$, but not for $\frac{1}{4}<\beta<\frac{1}{2}$) as $K\rightarrow \infty$ \citep{liu2020cauchy,fang2021heavy}. 

In this paper, we revisit methods of the first category, evaluate their asymptotic efficiencies, assess finite-sample numerical performance, and propose an ensemble method combining two complementary top performers for general applications. Despite increasing applications of complex and big data that require methods designed for the second and third categories, methods of the first meta-analytic scenario remain in high demand and present new challenges in many applications. Method development of the first category before the 1970-80s focuses on a class of methods aggregating transformed scores from the $p$-values: $T=\sum_{i=1}^K g(p_i)= \sum_{i=1}^K F_U^{-1}(p_i)$, where $F^{-1}_U(\cdot)$ is the inverse CDF of $U$. For example, $U$ is chi-squared distribution for Fisher test and standard normal distribution for Stouffer test.
\cite{littell1973asymptotic} showed that Fisher’s method is asymptotically optimal in terms of Bahadur relative efficiency, providing theoretical justification of the $\log$-transformation over the other types of transformations (see Section \ref{sec:ABOsection} for more details).

Despite optimal asymptotic efficiency of Fisher test, its finite-sample performance in terms of statistical power can be poor if only part of the $K$ $p$-values have signals. In this commonly encountered situation with heterogeneous signals, modified Fisher methods have been developed in the past ten years to adaptively aggregate signals only for the top ordered (i.e., the most significant) $p$-values: $T=-2\sum_{i=1}^m \log p_{(i)}$, where $p_{(i)}$ is the $i$-th ordered $p$-value and $m$ is data-driven, usually determined from an optimization criterion. \cite{li2011adaptively} proposed an adaptive Fisher procedure using partial sum optimized by the corresponding \underline{$p$}-values (denoted by AFp hereafter): $ T_{\text{AFp}}=\max_{1\leqslant j\leqslant K}-\log(h(\vec{p}, j))$, where $h(\vec{p}, j)=\bar{F}_{\chi^2_{2j}}(-2\sum_{i=1}^j\log p_{(i)})$ is the corresponding $p$-value of the partial sum, and $\bar{F}_{\chi_{2j}^2}(t)=1-F_{\chi_{2j}^2}(t)$, where $F_{\chi_{2j}^2}(t)$ denotes the CDF of chi-squared random variable with degrees of freedom $2j$.  Later, \cite{song2016screening} developed another adaptive Fisher procedure using partial sum optimized by \underline{$z$}-standardization similar to higher criticism (denoted by AFz  hereafter): $ T_{\text{AFz}}=\max_{1\leqslant j\leqslant K}\frac{-\sum_{i=1}^j \log p_{(i)}-\sum_{i=1}^{j} w(j, i)}{\sqrt{\sum_{i=1}^{j} w^{2}(j, i)}} \text { where } w(j, i)=\min \{1, i / j\}$. Another related strategy in the literature is to directly filter out $p$-values greater than a user-specified threshold $\tau\in (0,1]$. For example, the truncated Fisher with hard-thresholding (denoted by TFhard) $T_{\text{TFhard}}(\tau)=\sum_{i=1}^K-\log (p_i)\I_{\{p_i\leq \tau\}} $ \citep{zaykin2002truncated}, where $\I_{\{\cdot\}}$ denotes the indicator function. \cite{zhang2020tfisher} proposed truncated Fisher with soft-thresholding (TFsoft) to improve TFhard, arguing that the continuous soft-thresholding scheme can lead to more stable performance with varying input $p$-values. In both TFsoft and TFhard, the choice of $\tau$ is not straightforward. \cite{zhang2020tfisher} investigated the optimal choice of $\tau$ for TFhard under a theoretical setting of Gaussian mixture, where mixture probability and mean of the signals are known and $K\rightarrow\infty$.  However, such prior information is generally unknown in applications. To this end, they replaced a single user-specified $\tau$ with a user-specified set of thresholds $\mathcal{T}$ and proposed two omnibus tests for TFhard and TFsoft, which alleviate the issue of choosing $\tau$ to some extent but the selection of $\mathcal{T}$ is still prespecified and ad hoc. 

Notwithstanding the active development of modified Fisher methods, there is a lack of comprehensive and systematic evaluation of the asymptotic properties and finite-sample numerical performance of the methods in the first category. Our paper sets out to fill this gap. In Section \ref{sec:ABOsection}, we examine asymptotic Bahadur optimality (ABO) of six methods in the first category: Fisher, Stouffer, AFp, AFz, TFhard, and TFsoft. The two adaptive Fisher methods, AFp and AFz, provide estimates of the subset of $p$-values contributing to the signal. Therefore, we also investigate whether the estimates in these two methods consistently select the subset of $p$-values containing true signals (signal selection consistency). For completeness, we also examine asymptotic efficiencies for methods developed for sparse signals, including Cauchy, Pareto family, minimum $p$-value (minP), BJ, and HC. In Section \ref{sec:simulations}, we perform finite-sample numerical evaluations to compare statistical power of these methods under different $K$, signal strength, and proportions of true signals. Results of Sections \ref{sec:ABOsection} and \ref{sec:simulations} conclude complementary advantages of two top performers -- Fisher and AFp --, especially in varying proportions of true signals. Consequently, we develop a Fisher ensemble (FE) method in Section \ref{sec:FisherEnsemble} that applies a truncated Cauchy ensemble approach to combine Fisher and AFp. We prove asymptotic Bahadur optimality of FE (Section \ref{subsec:AggRV}) and demonstrate its consistently high performance in varying simulation scenarios (Section \ref{subsec:ensembleSimulation}). Section \ref{sec:Onesided} develops an extension of the FE method, namely FE$_{\text{CS}}$, for enhanced statistical power on detecting signals with concordant effect size directions. 
Section \ref{sec:AGEMAP} applies FE and FE$_{\text{CS}}$ as well as existing methods to a transcriptomic meta-analysis on biomarker and pathway detection for aging \citep{zahn2007agemap}. Section \ref{sec:discussion} provides final discussion and conclusion.
\section{Asymptotic efficiencies of existing methods}\label{sec:ABOsection}
This section investigates the asymptotic efficiencies of existing $p$-value combination methods. Since our focus is for the scenarios with finite-number signals, we slightly generalize the setup proposed in \cite{littell1973asymptotic} {(differences are discussed in Remark \ref{remark:diff})}, which uses the criterion of exact Bahadur relative efficiency \citep{bahadur1967rates}. Under this setting, Fisher's method is asymptotically Bahadur optimal \citep{littell1973asymptotic} and shows theoretical advantages of $\log$-transformation. Multiple modified Fisher's methods (AFp, AFz, TFhard, and TFsoft) have been developed to improve finite-sample statistical power, but their asymptotic efficiencies have not been investigated. Section \ref{subsec:ABOdef} introduces the problem setting and defines the exact slope of a hypothesis test, which is a natural concept derived from the exact Bahadur relative efficiency. 
Section \ref{subsec:MFisherABO} presents asymptotic Bahadur optimality (ABO) results of the four modified Fisher's methods. 

\subsection{Bahadur relative efficiency and exact slope}\label{subsec:ABOdef}
We first introduce the concept of exact slope of a hypothesis test \citep{bahadur1967rates}. Consider $(x_{1}, x_{2}, \cdots)$ an infinite sequence of independent observations of a random variable $X$ from probability distribution $P_\theta$ with parameter $\theta\in \Theta$. Let $T_n$ be a real-valued and continuous test statistic depending on the first $n$ observations $(x_1,\ldots, x_n)$, where large values of $T_n$ are considered to reject the null hypothesis. Denote $\PR_0(T_n<t)=\PR_\theta(T_n<t)$ for $\theta \in \Theta_0$. Further denote $p^{(n)}=1-F_n(t_n)$ as the $p$-value of observed $T_n=t_n$, where $F_n(t)=\PR_0(T_n<t)$. We then define the exact slope of $T_n$ as follows.

\begin{definition}\label{exactSlopeDef}
For the test statistic $T_n$ with $p$-value $p^{(n)}$, if there is a positive valued function $c(\theta)$, such that for any $\theta\in \Theta-\Theta_0$, 
$
-\frac{2}{n}\log p^{(n)}\rightarrow c(\theta) \text{ as $n\rightarrow \infty$}
$
with probability one. Then $c(\theta)$ is called the \texttt{exact slope} of $T_n$.
\end{definition}

As a simple example, consider testing for zero mean ($\mu=0$) with known variance under univariate Gaussian distribution and $T_n$ is the conventional $z$-test. It is easily seen that $c(\mu)=\mu^2$ is the exact slope of the $z$-test. Exact slope of a test naturally connects to the exact Bahadur efficiency between test statistics. Consider two sequences of test statistics $\{T_{n}^{(1)}\}$ and $\{T_{n}^{(2)}\}$ testing the same null hypothesis with exact slopes $c_1(\theta)$ and $c_2(\theta)$ respectively. We define the ratio $\phi_{12}(\theta)=c_1(\theta)/c_2(\theta)$ as the \textit{exact Bahadur relative efficiency} of $\{T_{n}^{(1)}\}$ relative to $\{T_{n}^{(2)}\}$, which compares the relative asymptotic efficiency between two test statistics. Indeed, considering any significance level $\alpha>0$, for $i=1,2$, denote $N^{(i)}(\alpha)$ as the smallest sample size such that, for any $n\geqslant N^{(i)}(\alpha)$, the $p$-value of $T^{(i)}_n$ is smaller than $\alpha$, one can show $
\lim_{\alpha \rightarrow 0} {N^{(2)}(\alpha)}/{N^{(1)}(\alpha)}=\phi_{12}(\theta),
$
which asymptotically characterizes the ratio of the smallest sample sizes of two test statistics required to attain the same sufficiently small significant level $\alpha$ \citep{littell1973asymptotic}.

For $\theta \in \Theta_0$, the $p$-value $p^{(n)}$ follows uniform distribution $\Unif(0,1)$. Lemma \ref{lemma1} shows the analogous ``exact slope" $-(2/n)\log p^{(n)}$ converges to zero with probability one. 
\begin{lemma}\label{lemma1}
For $\theta\in \Theta_0$, as $n$ diverges,
$
-(2/n)\log p^{(n)}\rightarrow 0 
$
with probability one.
\end{lemma}

In this paper, we extend the definition of exact slope to the null parameter space, where $c(\theta)=0$ for $\theta \in \Theta_0$. 

To benchmark the asymptotic efficiency of a $p$-value combination method, we then introduce the theoretical setup adopted from the framework in \cite{littell1973asymptotic}. Suppose we have $K<\infty$ sequences of test statistics $\{T_{n_1}^{(1)}\},\ldots,\{T_{n_K}^{(K)}\}$ for testing $\theta_i\in \Theta_0^{(i)}$ for $1\leq i\leq K$. Assume for all the sample sizes $n_1,\ldots,n_K$, and when $\theta_i\in \Theta_0^{(i)}$ for $1\leq i\leq K$,  $\{T_{n_1}^{(1)}\},\ldots,\{T_{n_K}^{(K)}\}$ are independently distributed. Denote $p_i^{(n_i)}$ as the $p$-value of the $i$-th test statistic $T_{n_i}^{(i)}$. For each $1\leq i\leq K$, assume that the sequence $\{T_{n_i}^{(i)}\}$ has exact slope $c_i(\theta_i)$ as 
$
-(2/n_i)\log p_i^{(n_i)}\rightarrow c_i(\theta_i)\geqslant 0
$
with probability one as $n_i\rightarrow \infty$. We further assume the sample sizes $n_1,\ldots,n_K$ satisfy $n=(1/K)\sum_i^K n_i$ and $
\lim_{n\rightarrow\infty} (n_i/n)=\lambda_i$, where $\lambda_i>0$ and $\sum_i^K \lambda_i=K$. Under the above setup, the goal of any $p$-value combination method is to test 
\begin{align}
H_0: \cap_{i=1}^K \big\{\theta_i \in \Theta^{(i)}_0\big\} \text{ versus } H_1:  \cup_{i=1}^K \big\{\theta_i \in\Theta^{(i)}-\Theta^{(i)}_0\big\}.\label{goal1}
\end{align}
For simplicity, we assume under the null
$$\lambda_{1} c_{1}\left(\theta_{1}\right) \geqslant \lambda_{2} c_{2}\left(\theta_{2}\right) \geqslant \ldots \geqslant \lambda_{K} c_{K}\left(\theta_{K}\right) \geqslant 0,$$
where the first $\ell$ $p$-values have true signals (i.e., $\theta_i$'s belong to $\Theta^{(i)}-\Theta_0^{(i)}$ for $1\leq i\leq\ell$) with exact slopes $c_i(\theta_i)>0$, while $c_i(\theta_i)=0$ for the remaining $\theta_i\in \Theta^{(i)}_0$, $\ell +1\leq i\leq K$.  
\begin{remark}\label{remark:diff}
There are two differences between the original setup in \cite{littell1973asymptotic} and ours. First, \cite{littell1973asymptotic} assumed that all the studies have strictly positive exact slopes, while we allow some studies to have zero-valued exact slopes. Second, \cite{littell1973asymptotic} considered a general parameter space $\Theta$ while we consider a product of parameter spaces $\Theta^{(1)}\times\Theta^{(2)}\times\cdots\times\Theta^{(K)}$. Although differences exist, one can still establish the results in \cite{littell1973asymptotic} by combining their arguments with Lemma \ref{lemma1}.
\end{remark}
Following Theorem 2 and arguments in Section 4 in \cite{littell1973asymptotic}, under the alternatives, the maximum attainable exact slope for any $p$-value combination method is $\sum_{i=1}^\ell \lambda_ic_i(\theta_i)$. Hence we define the asymptotic Bahadur optimality (ABO) of a $p$-value combination method as follows. 
\begin{definition}\label{ABODef}
Denote $\vec{\theta}=(\theta_1,\ldots,\theta_K)$. Under the above setup, a $p$-value combination test $H(p_1,\ldots,p_K)$ is asymptotically Bahadur optimal (ABO) if its exact slope $C_H(\vec{\theta})$ satisfies
$
C_H(\vec{\theta})=\sum_{i=1}^\ell\lambda_ic_i(\theta_i).
$
\end{definition}

\subsection{Asymptotic Bahadur optimality property of $p$-value combination methods}\label{subsec:MFisherABO}

\cite{littell1973asymptotic} showed that Fisher test is ABO while Stouffer and minP tests are not. Except for these methods, there is a lack of asymptotic efficiency analysis of the other methods. This subsection focuses on discussing four modified Fisher methods: AFp, AFz, TFhard, and TFsoft. We additionally analyze four methods designed for combining sparse and weak signals: Cauchy, Pareto, BJ, and HC. As expected, the latter four tests do not enjoy ABO property, and the proofs are outlined in the supplement. The theoretical results of ABO, exact slope, and signal selection consistency (to be discussed in Theorem \ref{AFpConsistency} and remarks S3, S5, and S6) are summarized in Table \ref{tab:ABOTable}.

    \begin{table}[!htp]

     \centering
  
    \caption{Results of asymptotic properties of 11 $p$-value combination methods: Fisher, Stouffer, four modified Fisher (AFp, AFz, TFhard and TFsoft) and five methods designed for sparse and weak signal (Cauchy, Pareto, minP, BJ and HC).}    \label{tab:ABOTable}
    
\begin{tabular}[t]{l|l|l|p{25mm}|l}
\bottomrule
Methods& ABO & Exact slopes & Signal selection consistency& Proofs \\
\toprule
Fisher & Yes& $\sum_{i=1}^\ell\lambda_ic_i(\theta_i)$&-- & Theorem S1\\
Stouffer & No  & $\frac{1}{K}\left[\sum_{i=1}^\ell (\lambda_ic_i(\theta_i))^\frac{1}{2}\right]^2$&-- & Theorem S1\\
AFp     & Yes& $\sum_{i=1}^\ell\lambda_ic_i(\theta_i)$  & Yes & Theorems \ref{AFpABO} \& \ref{AFpConsistency}\\
AFz     & No &$\leqslant \max_j\frac{\sum_{i=1}^{j}\lambda_ic_i(\theta_i)}{\sqrt{\sum_{i=1}^{j}\min^2\{1,i/j\}}}$& No & Theorem \ref{AFzABO} \\
TFhard & Yes & $\sum_{i=1}^\ell\lambda_ic_i(\theta_i)$&-- &Theorem \ref{TFthm}\\
TFsoft & Yes & $\sum_{i=1}^\ell\lambda_ic_i(\theta_i)$&-- &Theorem \ref{TFthm}\\
Pareto & No &$\max _{i} \lambda_{i} c_{i}\left(\theta_{i}\right)$ &--& Theorem S2\\
Cauchy   & No&$\max _{i} \lambda_{i} c_{i}\left(\theta_{i}\right)$ & -- &Theorem S3\\
minP & No & $\max _{i} \lambda_{i} c_{i}\left(\theta_{i}\right)$&--& \cite{littell1973asymptotic}\\

BJ     & No &$[\frac{1}{K}\max _{i} i \lambda_{i} c_{i}\left(\theta_{i}\right),\max _{i} i \lambda_{i} c_{i}\left(\theta_{i}\right)]$& No & Theorem S4\\
HC     & No &--   & No & Proposition S1\\
\toprule
\end{tabular}
\end{table}

Recall that Fisher and the four modified Fisher methods combine $p$-values using the following test statistics: 
\begin{align*}
    &T_{\text{Fisher}}=\sum_{i=1}^K -2\log p_{(i)};\;
    T_{\text{AFz}}=\max_{1\leqslant j\leqslant K}\frac{-\sum_{i=1}^j \log p_{(i)}-\sum_{i=1}^{K} w(i, j)}{({\sum_{i=1}^{K} w^{2}(i, j)})^{\frac{1}{2}}};\\
    &T_{\text{AFp}}=\max_{1\leqslant j\leqslant K}-\log(\bar{F}_{\chi^2_{2j}}(-2\sum_{i=1}^j\log p_{(i)}))
    ;\;T_{\text{TFhard}}(\tau)=\sum_{i=1}^K (-2\log p_i)\I_{\{p_i\leqslant \tau\}};\\
    &T_{\text{TFsoft}}(\tau)=\sum_{i=1}^K (-2\log p_i+2\log\tau)_+.
\end{align*}
Here $w(i, j)=\min \{1, j / i\}$. In addition, $\tau\in (0,1]$ is a user-specified constant for the two truncated Fisher methods and $(x)_+$ denotes $\max(x,0)$.
 
All the five methods can be characterized in the form of $H(-\log p_1,\ldots,-\log p_K)$ by some function $H(\cdot)$. With the $\log$-transform on $p$-values as a key ingredient, the above methods potentially can achieve high asymptotic efficiency. Indeed, combining with Lemma \ref{lemma1}, by using almost the same arguments in \cite{littell1973asymptotic}, one can show that Fisher test attains ABO, presented in Theorem S1 for completeness. 

Although achieving high asymptotic efficiency, the Fisher test has been shown to have poor performance empirically when only small part of $p$-values contain signals (e.g., 2 out of 10 $p$-values have signals); see \cite{song2016screening} and \cite{li2011adaptively} for more discussions. Many modified Fisher methods have been proposed to address this problem \citep{zaykin2002truncated,yu2009pathway,kuo2011novel,zhang2020tfisher,li2011adaptively,song2016screening}. The idea is to filter out large $p$-values that are less likely to carry signals and reduce the impact of noise, while still using the $\log$-transformation on $p$-values to achieve high efficiency. Particularly, AFp and AFz combine the first $m$ smallest ordered $p$-values. Both methods use some optimization criterion that adaptively selects $m$ to achieve superior finite-sample power in varying proportions of signals. Whether AFp and AFz retain the ABO property of Fisher is an intriguing question and is investigated below. In fact, we will surprisingly find in the following two theorems that AFp is ABO, but AFz is not.

\begin{theorem}[AFp is ABO]\label{AFpABO}
 Under the setup in Section \ref{subsec:ABOdef}, $T_{\text{AFp}}$ is similar to Fisher test to be ABO with exact slope
$C_{\text{AFp}}(\vec{\theta})=\sum_{i=1}^\ell\lambda_ic_i(\theta_i).$
\end{theorem} 

\begin{theorem}[AFz is not ABO]\label{AFzABO}
Under the setup in Section \ref{subsec:ABOdef}, consider the following test statistic
$
T_\text{A}=\max_{1\leqslant j\leqslant K} \frac{-2\sum_{i=1}^j \log p_{(i)}-A_j}{B_j},
$
where $B_j>0$ and $A_j$ are some finite constants only depend on j and K. Assume there is no tie for $\frac{\sum_{i=1}^j\lambda_ic_i(\theta_i)}{B_j}$, j=1,....,K, and $B_j$ is monotonic increasing. Then $T_A$ is not ABO with exact slope
$
C_{\text{A}}(\vec{\theta})\leqslant \max_{1\leqslant j\leqslant \ell} {(B_1/B_j)\sum_{i=1}^j\lambda_ic_i(\theta_i)}.
$
The equality holds if and only if $\ell=1$ (i.e., there is only one signal inside the $K$ $p$-values). 
\end{theorem}
By taking $A_j=2\sum_{i=1}^{K} w(i, j)$ and $B_j=2(\sum_{i=1}^{K} w^2(i, j))^{\frac{1}{2}}$, $T_A$ reduces to $T_{\text{AFz}}$, indicating that AFz is not ABO in general (e.g., a special case that AFz is ABO is when $\ell=1$). 

AFp's better asymptotic efficiency property compared to AFz may be due to its attempt to estimate the subset of $p$-values with true signals. Consider the equivalent form of AFp for combining independent $p$-values:
$$
T_{\text{AFp}}^\prime=\min_{\vec{{w}}} \bar{F}_{\chi^2_{2 (\sum_{i=1}^{K} w_{i})}}(-2 \sum_{i=1}^{K} w_{i} \log p_{i}),
$$
where $\vec{w}=\left(w_{1}, \ldots, w_{K}\right) \in\{0,1\}^{K}$ is the vector of binary weights that identify the candidate subset of $p$-values with true signals. Note that $T_{\text{AFp}}^\prime$ is the original form proposed in \cite{li2011adaptively}. Denote by 
$
\hat{\vec{w}}=\argmin_{\vec{w}} \bar{F}_{\chi^2_{2 (\sum_{i=1}^{K} w_{i})}}(-2 \sum_{i=1}^{K} w_{i} \log p_{i})
$
and let $\vec{w}^{*}=\left\{(w_1^*,\cdots , w_K^*): w_{k}^*=1 \text { if } \theta_{i} \in \Theta-\Theta_0 \text { and } w_{k}^*=0 \text { if } \theta_{i}\in \Theta_0\right\} $ be the indicators of the true signals 
We can show signal selection consistency of AFp in the following theorem.
\begin{theorem}[signal selection by AFp is consistent]\label{AFpConsistency}
Under the setup in Section \ref{subsec:ABOdef}, $\hat{\vec{w}}\rightarrow \vec{w}^{*}$ as $n\rightarrow \infty$ in probability for the AFp test. 
\end{theorem}
The following Theorem \ref{TFthm} states that for any given value of $\tau\in (0,1]$, TFhard and TFsoft are ABO:
\begin{theorem}[TFhard and TFsoft are ABO]\label{TFthm}
Under the setup in Section \ref{subsec:ABOdef}, TFhard and TFsoft are ABO with exact slopes
$
C_{\text{TFhard}}(\vec{\theta})=C_{\text{TFsoft}}(\vec{\theta})=\sum_{i=1}^\ell\lambda_ic_i(\theta_i).
$
\end{theorem}
Although TFhard and TFsoft are ABO, the choice of $\tau$ may significantly impact their finite-sample performance \citep{zhang2020tfisher}. To address this issue, \cite{zhang2020tfisher} proposed the following omnibus tests for both methods (denoted by oTFhard and oTFsoft, respectively):
\begin{align*}
    T_{\text{oTFhard}}=\min_{\tau\in \mathcal{T}} 1-F_{U_{\text{TFhard}}(\tau)}(T_{\text{TFhard}}(\tau));\;
    T_{\text{oTFsoft}}=\min_{\tau\in \mathcal{T}} 1-F_{U_{\text{TFsoft}}(\tau)}(T_{\text{TFsoft}}(\tau)),
\end{align*}
where $\mathcal{T}=\{\tau_1,\ldots,\tau_m\}$ is a user-specified set of the candidates of $\tau$. Here $U_{\text{TFhard}}(\tau)$ and $U_{\text{TFsoft}}(\tau)$ denote the random variables that follow the null distributions of $T_{\text{TFhard}}(\tau)$ and $T_{\text{TFsoft}}(\tau)$, respectively.
Although the omnibus tests alleviate the issue of sensitivity of the choice of $\tau$ for both TFhard and TFsoft to some extent, selection of $\mathcal{T}$ is still user-specified and subjective. In addition, \cite{zhang2020tfisher} derive the null distributions of both omnibus tests in an asymptotic sense as $K\rightarrow \infty$, which may not be accurate for small $K$ with small $p$-value thresholds that are commonly used in applications, such as genomics studies, to handle multiplicity.

Proofs of theorems for Fisher and modified Fisher methods in this subsection can be found in Supplement Section S2.1. For completeness, we also show that methods designed for combining sparse and weak signals, such as Cauchy, Pareto, BJ and HC, are not ABO (Supplement Section S1) and the proofs are available in Supplement Section S2.3. In conclusion, Fisher, AFp, TFhard and TFsoft are the only four methods with ABO property. AFp and AFz are two methods to provide signal selection (i.e., subset estimation of the true signal) and AFp is the only method to have consistency in the signal identification.

\section{Power comparison in finite-sample simulations}\label{sec:simulations}
Although Section \ref{sec:ABOsection} evaluates asymptotic efficiencies of $p$-value combination methods, the finite-sample statistical power of the methods under different proportions of signals has not been assessed. In this section, we evaluate six methods that are designed for relatively frequent signal setting described in Section \ref{sec:ABOsection}: Fisher, Stouffer, AFp, AFz, TFhard, and TFsoft. Additionally, we also evaluate methods designed for combining sparse and weak signals for completeness: minimum $p$-value (minP), Cauchy (CA), harmonic mean (HM), Berk \& Jongs (BJ), and higher criticism (HC). As TFhard and TFsoft are sensitive to the choice of tuning parameter $\tau$, for a fair comparison, we use the corresponding omnibus tests, oTFhard and oTFsoft, instead. The tuning candidate set $\mathcal{T}$ is set to be $\{0.01,0.05,0.5,1\}$, which is used in the original paper \citep{zhang2020tfisher}. 

For better illustration, we first present results of the six methods designed for combining relatively frequent signals in Figure \ref{power1}. Results comparing all eleven methods can be found in Supplement Figure S1, where modified Fisher's methods generally dominate other methods designed for sparse and weak signals unless the signals are indeed sparse and weak (e.g., only $\ell=1$ true signal out of $K=10$ $p$-values). However, in such cases, methods such as AFp and AFz still have comparable power with the top-performing methods such as minP.   

We simulate $X=(X_1,\ldots,X_K)\overset{D}{\sim} N(\vec{\mu},I_K)$, where  $\vec{\mu}=(\mu_1,\mu_2,\ldots,\mu_K)$ contains $\ell$ non-zero signals $\mu_1=\cdots=\mu_\ell=\mu_0$ and $K-\ell$ with no signal ($\mu_{\ell+1}=\cdots=\mu_K=0$). We evaluate for a wide range of $K$ and proportions of true signals $\ell/K$: $K=10,20,40,80$, and $\ell/K=0.1,0.2,\ldots,0.7$. We also vary $\mu_0=0.5,0.65,\ldots,5$ for a broad range of signal strength. The $p$-values are calculated by two-sided test $p_i=2(1-\Phi(\left|X_i\right|))$ for $i=1,\ldots,K$. For each combination of parameter values, we draw $10^6$ Monte Carlo samples to calculate the critical values for all the methods at significance level $\alpha=0.01$, Since the $p$-value calculation algorithms for some methods, such as oTFsoft and oTFhard, are not accurate for small $K$.

Figure \ref{power1} shows the empirical power of Fisher, Stouffer, and four modified Fisher methods at $\alpha=0.01$. For a given $K$ and proportion of signals $\ell/K$, we choose the smallest $\mu_0$ such that the best method has at least 0.5 statistical power, which allows optimized visualization and comparison of different methods in different signal settings.
We first note that AFz is inferior to the other modified Fisher methods, consistent with our theoretical result that AFz is not ABO. We further note that AFp, oTFhard, and oTFsoft have comparable performance across varying proportions of signals, significantly dominating Fisher for detecting infrequent signals (e.g., when the proportion of true signals is smaller than 0.3). On the contrary, Fisher outperforms all other methods for detecting frequent signals (e.g., when the proportion of true signals is greater than 0.5). In many real applications (e.g., the transcriptomic meta-analysis in Section \ref{sec:AGEMAP}), the $p$-value combination test is repeated many times (i.e., for each gene). It is expected that some true biomarkers are more homogeneous with frequent true signals and some with less frequent signals. The result in Figure \ref{power1} shows the need to develop an ensemble method to integrate the advantages of Fisher and one of the top-performing modified Fisher methods, which is presented in the next section.   

\begin{figure}
\centering
\includegraphics[scale=1]{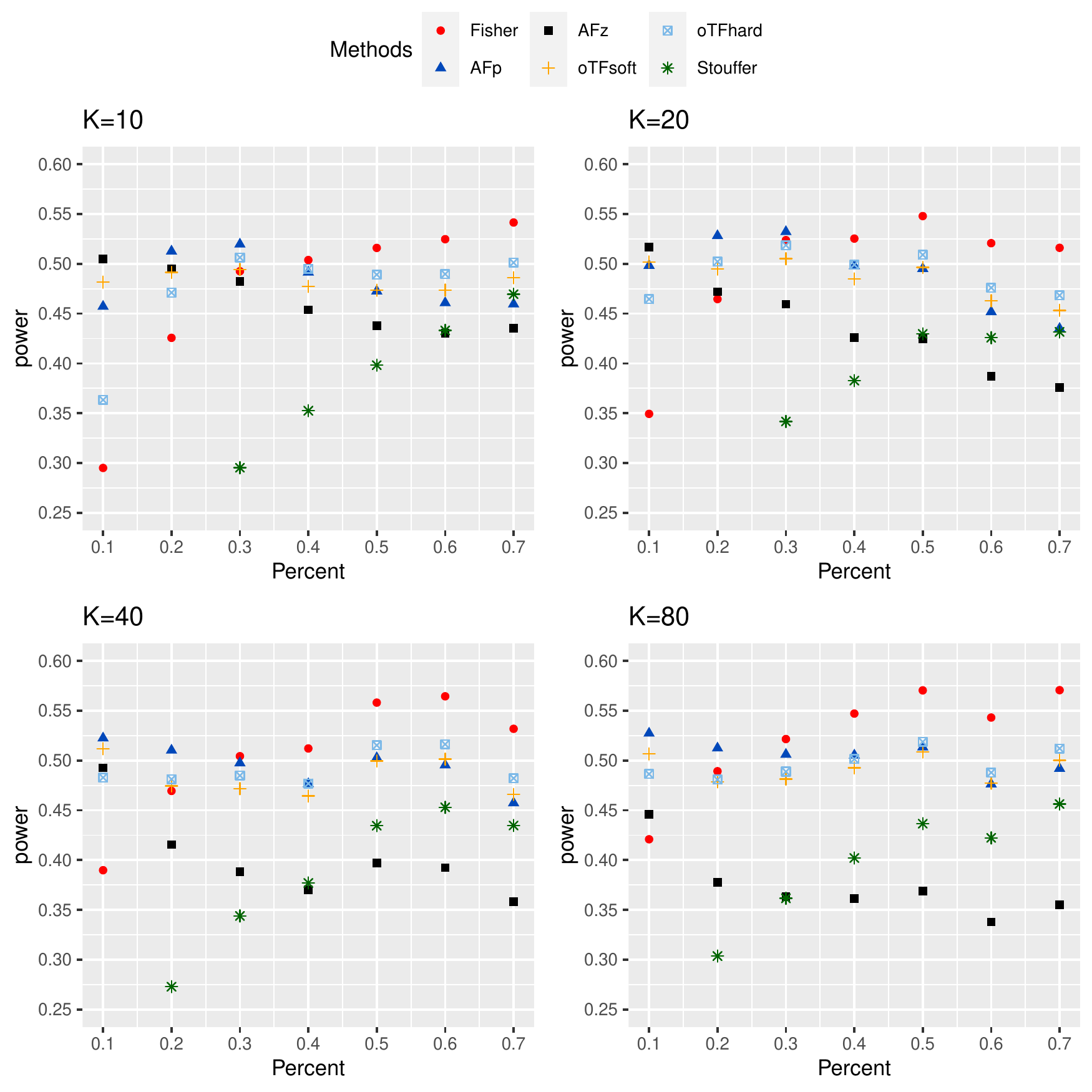}
\caption{Statistical power of Fisher, Stouffer, and four modified Fisher's methods at significance level $\alpha=0.01$ across varying frequencies of signals $\ell/K=0.1,0.2,\ldots,0.7$ and varying numbers of combined $p$-values $K=10,20,40,80$. 
The standard errors are negligible compared to the scale of the mean power (smaller than $0.1\%$ of the power) and hence omitted.  
The results of Stouffer with power smaller than 0.25 when $\ell/K=0.1$ or $0.2$ are omitted. 
}
\label{power1}
\end{figure}

\section{Fisher ensemble to combine Fisher and AFp }\label{sec:FisherEnsemble}
As shown in Sections \ref{sec:ABOsection} and \ref{sec:simulations}, Fisher and three modified Fisher methods (AFp, TFhard, and TFsoft) are ABO, and have complementary strength in finite-sample evaluation of varying proportions of true signals. A natural idea is to ensemble Fisher and one of the three modified Fisher methods for more stable and universally competitive performance. Since oTFhard and oTFsoft methods require an ad hoc decision of user-specified set $\mathcal{T}$ and their existing computing algorithms are not accurate for small $K$, we choose to develop an ensemble method to combine Fisher and AFp in this section. In Section \ref{subsec:truncatedCA}, we propose an ensemble approach, namely Fisher ensemble (FE), using truncated Cauchy method \citep{fang2021heavy} to combine Fisher and AFp. In section \ref{subsec:AggRV}, we provide theoretical support of FE and show that FE is ABO. Section \ref{subsec:ensembleSimulation} presents simulation results similar to Section \ref{sec:simulations} to demonstrate the balanced and superior performance of FE across varying proportions of true signals. 

\subsection{Fisher ensemble by truncated Cauchy integration}\label{subsec:truncatedCA}
Denote by $p^{\text{Fisher}}$ and $p^{\text{AFp}}$ the $p$-values derived from Fisher and AFp combination tests, respectively. We propose to ensemble the two methods by combining their $p$-values using $T_{h}=[h(p^{\text{Fisher}})+ h(p^{\text{AFp}})]/2$ with function $h$. {Since $p^{\text{Fisher}}$ and $p^{\text{AFp}}$ can be highly dependent, one option is to use the Cauchy combination test with $h(p)=\tan(\pi(\frac{1}{2}-p))$, as theorems and simulations in \cite{liu2020cauchy} and \cite{liu2019acat} show that the Cauchy combination test is robust to dependency of combined $p$-values, and further results in a fast-computing algorithm with Cauchy distribution under the null hypothesis (i.e., the null distribution is standard Cauchy)}.
This Cauchy ensemble approach is, however, problematic when either $p^{\text{Fisher}}$ or $p^{\text{AFp}}$ is close to 1. In this case, the Cauchy transformation generates a $-\infty$ score and the power is greatly reduced. We propose to adopt the truncated Cauchy transformation \citep{fang2021heavy} in our Fisher ensemble (FE) method by 
\begin{align}
T_{\text{FE}}=[h_\delta(p^{\text{Fisher}})+ h_\delta(p^{\text{AFp}})]/2,
\end{align}\label{eq:FE}
where $h_\delta(p)=h(p)\I_{\{p<1-\delta\}}+h(1-\delta)\I_{\{p\geqslant 1-\delta\}}$ and $\delta$ is set as 0.01 as justified in the original paper. The purpose of using $h_{\delta}(p)$ instead of $h(p)$ is to avoid the large negative score issue of the transformation by Cauchy distribution; see \cite{fang2021heavy} for more details. Except for avoiding large negative score issue, ensemble by truncated Cauchy using $h_{\delta}(p)$ performs almost identically to Cauchy $h(p)$. Supplement Section S3.4 provides numeric examples where the ensemble method using $h_{\delta}(p)$ performs better than that using $h(p)$. 

In the implementation, FE is fully data-driven with fast-computing algorithms. Indeed, for $p_1,\ldots,p_K \overset{\text{i.i.d.}}{\sim} \Unif(0,1)$, null distribution of Fisher test follows chi-squared distribution with degrees of freedom $2K$. Fast $p$-value calculation for AFp has been previously developed in \cite{huo2020p} using importance sampling. Using the null reference distribution library for AFp pre-built by \cite{huo2020p}, one can calculate the $p$-value of AFp in linear computing time (R package ``AWFisher''). Finally, Proposition 3 in \cite{fang2021heavy} has shown that the truncated Cauchy approach proposed here has at most $(1+\delta)^2-1=(1+0.01)^2-1=2\%$ inflation of the type I error control if we naively use Cauchy distribution as the null distribution (see Proposition S2 in supplement). In other words, if the calculated $p$-value is 0.05 from null distribution using Cauchy distribution, the true $p$-value from null distribution using truncated Cauchy is between 0.05 and $0.05\cdot (1+0.01)^2=0.051$. As a result, fast $p$-value computation for Fisher ensemble $T_{\text{FE}}$ is warranted. Table S1 in Section S3.1 justifies the above fast-computing procedure, where we show the type I error control for FE with $\delta=0.01$ is accurate for $\alpha\leqslant 0.05$ across a broad range of $5\leq K\leq 100$.

\subsection{Asymptotic efficiency of Fisher ensemble}\label{subsec:AggRV}
In this subsection, we will show that Fisher ensemble (FE) is asymptotically Bahadur optimal (ABO). We first introduce a heavy-tailed distribution family, namely regularly-varying distribution $R$ \citep{mikosch1999regular}, where Cauchy and truncated Cauchy distributions are special cases of the family. Consider an ensemble method induced by a regularly-varying distribution (e.g., truncated Cauchy in our case) to combine multiple $p$-value combination methods (e.g., Fisher and AFp in our case). The ensemble method will be shown to be ABO if at least one of the $p$-value combination methods is ABO. Since both Fisher and AFp are ABO and truncated Cauchy is a regularly varying distribution, we conclude that Fisher ensemble is also ABO. Below, we outline the definition of regularly-varying distribution and the theorem. The detailed proof is available in Supplement Section S2.2.

\begin{definition}\label{regdist}
A distribution $F$ is said to belong to the regularly-varying tailed family with index $\gamma$ (denoted by $F\in R_{-\gamma}$) if
$\lim_{x\rightarrow\infty}\frac{\bar{F}(xy)}{\bar{F}(x)}=y^{-\gamma}$
for some $\gamma > 0$ and all $y>0$. 
\end{definition}
We denote the whole family of regularly varying tailed distributions by $R$.  For
two positive functions $u(\cdot)$ and $v(\cdot)$, we write $u(t)\sim v(t)$ if $\lim_{t\rightarrow \infty}\frac{u(t)}{v(t)} = 1$.
It can be shown that every distribution $F$ belonging to $R_{-\gamma}$ can be characterized
by $\bar{F}(t)\sim L(t)t^{-\gamma},$
where $\bar{F}(t)=1-F(t)$ and $L(t)$ is a slowly varying function. A function $L$ is called slowly varying if $\lim_{y\rightarrow\infty}\frac{L(ty)}{L(y)}=1$ for any $t>0$. Regularly varying distribution is a wide class of heavy-tailed distributions, which includes Cauchy, harmonic mean, Pareto and truncated Cauchy.

Consider $L<\infty$ $p$-value combination test statistics $T_1,\ldots,T_L$, each combining $\vec{p}=(p_1,\cdots,p_K)$. Denote by $p_{T_1},\ldots,p_{T_L}$ the resulting $p$-values of $T_1,\cdots,T_L$. In Fisher ensemble, we have $L=2$ and $(T_1, T_2)$ are Fisher and AFp. Under Definition \ref{regdist}, consider the following ensemble method by a regularly varying tailed distribution:
$$
T_{\text{RV}}(\gamma)=\sum_{i=1}^Lg_{\gamma}(p_{T_i})=\sum_{i=1}^LF_{U(\gamma)}^{-1}(1-p_{T_i}),
$$
where $F_{U(\gamma)}$ is CDF of $U(\gamma)$ and $U(\gamma)\in R_{-\gamma}$. Under the null hypothesis, the test statistic transforms all the $p_{T_i}$'s into regularly varying tailed random variables with index $\gamma$. The following theorem suggests that under mild conditions, the ensemble method by regularly varying tailed distribution has ABO property.

\begin{theorem}\label{regthm}
For each $i=1,\ldots,L$, let $C_i(\vec{\theta})$ be the exact slope of $T_i$ and assume $\max_{1\leqslant i\leqslant L}C_i(\vec{\theta})>0$. 
Let $C_{\text{RV}}^{(\gamma)}$ be the exact slope of $T_{\text{RV}}(\gamma)$, if one of the following two conditions holds: (C1) $F^{-1}_{U(\gamma)}(1-p)$ is bounded below: $F^{-1}_{U(\gamma)}(1-p)\geqslant \nu$ for some constant $\nu$ and $\forall p\in [0,1]$;
(C2) All the $T_i$'s have non-zero exact slopes: $\min_{1\leqslant i\leqslant L} C_i(\vec{\theta})>0$.
then we have
$
C_{\text{RV}}^{(\gamma)}(\vec{\theta})= \max_{1\leqslant i\leqslant L}C_i(\vec{\theta}).
$
\end{theorem}
\begin{remark}
As $h_{\delta}(p)$ (truncated Cauchy) is bounded below while $h(p)$ (Cauchy) is not, using $h_{\delta}(p)$ rather than $h(p)$ can satisfy Condition (C1) in Theorem \ref{regthm}. In general, if Condition (C1) is not satisfied, Condition (C2) is a mild condition (meaning all tests $T_i$ are at least minimally effective and have non-zero slope) but Condition (C2) is not always easy to check or satisfied in practice. For example, when we aggregate methods combining left one-sided $p$-values and right one-sided $p$-values in Section \ref{sec:FisherEnsemble}, methods only combining left one-sided $p$-values will produce $p$-values converging to one when there exist only positive effects. See Section \ref{sec:Onesided} and Supplement Section S3.4 for more details. Hence, it is easier to truncate a general regularly varying transformation to satisfy Condition (C1).
\end{remark}
Theorem \ref{regthm} suggests that $T_{\text{RV}}(\gamma)$ is ABO as long as at least one of $T_1,\ldots,T_L$ methods is ABO. Consequently, Fisher ensemble is ABO since truncated Cauchy belongs to regularly-varying tailed distribution and both Fisher and AFp are ABO. 

\subsection{Finite-sample power comparison of Fisher ensemble}\label{subsec:ensembleSimulation}
\begin{figure}
\centering
\includegraphics[scale=1]{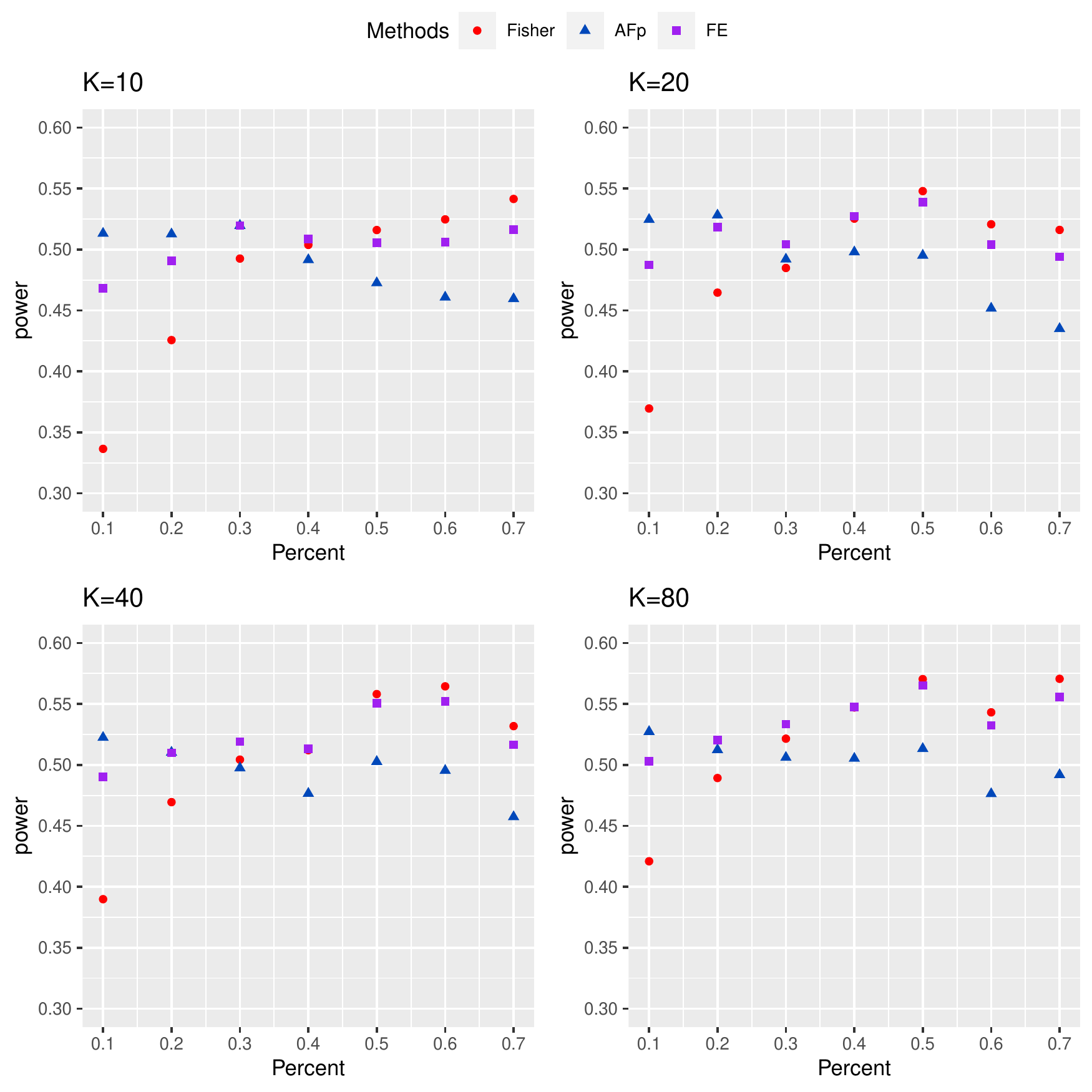}
\caption{Statistical power of FE, Fisher, and AFp at significance level $\alpha=0.01$ across varying frequencies of signals $\ell/K=0.1,0.2,\ldots,0.7$ and varying numbers of combined $p$-values $K=10,20,40,80$. 
The standard errors are negligible and hence omitted. 
}
\label{power2}
\end{figure}

In this subsection, we evaluate the finite-sample power of FE. To illustrate that FE can take advantages of integrated methods, we also include AFp and Fisher as the baseline methods. We use the same simulation scheme in Section \ref{sec:simulations} to generate the simulated data.   
Figure \ref{power2} shows statistical power of FE, AFp, and Fisher at $\alpha=0.01$. Similar to Figure \ref{power1}, For a given proportion of signals $\ell/K$ and number of combined $p$-values $K$, we choose the smallest $\mu_0$ that allows the best method to have power larger than 0.5. As expected, we note that FE has a stable statistical power that is comparable to the better of Fisher and AFp in different proportions of true signals. Specifically, when the proportion of signals is high, FE performs close to Fisher and is superior to AFp. When the proportion is low, FE performs close to AFp and outperforms Fisher. In Supplement Figure S2, we implement another Fisher ensemble method combining Fisher, AFp, and minP. As expected, its power for only one signal (i.e., $\ell=1$) is improved but at the expense of a large reduction of power when signals are frequent. From the asymptotic efficiency in Section \ref{subsec:AggRV} and simulations above, we recommend using the Fisher ensemble method combining Fisher and AFp for general applications.
\section{Detection of signals with concordant directions}\label{sec:Onesided}
\subsection{Fisher ensemble focused on concordant signals (FE$_{\text{CS}}$)}
For all methods we have discussed so far, the global hypothesis setting is designed for two-sided tests, regardless of directions of the effects. Recall from Equation \ref{goal1} that the hypothesis testing considered is $H_0: \cap_{i=1}^K \left\{\theta_i=0\right\} \text{ vs } H_1:  \cup_{i=1}^K \left\{\theta_i \neq 0\right\}$. Consider the alternative hypothesis that only the first $\ell$ $p$-values have true signals (i.e., $\theta_i\neq 0$ for $1\leq i\leq\ell$ and $\theta_{\ell+1}=\cdots=\theta_K=0$). The two-sided tests to obtain $p_i$ ($1\leq i\leq K$) cannot guarantee signals with concordant directions ($\text{sgn}(\theta_1)=\cdots=\text{sgn}(\theta_{\ell})$, denoted by $\text{sgn}(\cdot)$ the sign function), which is desirable in most applications. For example, when conducting meta-analysis of $K$ transcriptomic studies believed to be relatively homogeneous, we are interested in identifying biomarkers concordantly up-regulated or down-regulated. For this problem, \cite{owen2009karl} revisited the Pearson test statistic and proposed to use
$
   T_{\text{Pearson}} =\min\{\tilde{p}^{\text{\text{Fisher}},L},\tilde{p}^{\text{\text{Fisher}},R}\},
$
where $\tilde{p}^{Fisher,L}$ and $\tilde{p}^{Fisher,R}$ uses Fisher to combine the left and right one-sided $p$-values respectively, and the Pearson test takes the more significant one as the test statistic. In this subsection, we similarly extend the Fisher ensemble method to use truncated Cauchy approach to combine the two left and right one-sided $p$-values of Fisher and AFp (denoted by FE$_{\text{CS}}$; \underline{F}isher \underline{e}nsemble for \underline{c}oncordant \underline{s}ignal): 
\begin{align*}
T_{\text{FE}_{\text{CS}}}&=(1/4)[h_{\delta}(\tilde{p}^{\text{Fisher},L})+h_{\delta}(\tilde{p}^{\text{Fisher},R})+h_{\delta}(\tilde{p}^{\text{AFp},L})+h_{\delta}(\tilde{p}^{\text{AFp},R})].
\end{align*}
\begin{remark}
When combining one-sided $p$-values, it is common to observe $p$-values close to 1 and it is critical to use truncated Cauchy, instead of Cauchy, to avoid $-\infty$ scores.   
\end{remark}
\begin{remark}
Let $C^{L}(\vec{\theta})$ be the maximum attainable exact slope for any $p$-value combination method combining left one-sided $p$-values, and define $C^{R}(\vec{\theta})$ in a similar manner for right one-sided $p$-values. By Theorem \ref{regthm}, the exact slope of FE$_{\text{CS}}$ is $\max\{C^{L}(\vec{\theta}),C^{R}(\vec{\theta})\}$, indicating high asymptotic efficiency 
as even if one has the prior knowledge of the effect size direction, it is impossible to design a $p$-value combination method with a larger exact slope for detecting concordant signals.
\end{remark}
For computation, similar to FE, one can use $p$-value calculation for standard Cauchy to calculate $p$-values for FE$_{\text{CS}}$ and there is at most $(1+0.01)^4-1=4\%$ inflation on type I error control using the standard Cauchy when $\delta=0.01$. This approximation procedure is justified by simulation results in Table S1 in Section S3.1 for a broad range of significance levels $\alpha$ and numbers of input $p$-values $K$.

\subsection{Finite-sample power comparison of Fisher ensemble for concordant signals}\label{subsec:concordantFE}
In this subsection, we evaluate the finite-sample power of FE$_{\text{CS}}$ with $\delta=0.01$. To demonstrate the advantages of FE$_{\text{CS}}$, we also include the regular FE and Pearson as the baseline methods. We use the same simulation scheme in Section \ref{sec:simulations} to generate the simulated data. For FE$_{\text{CS}}$ and Pearson, the one-sided $p$-values are generated by $\tilde{p}_i^{(L)}=1-\Phi(X_i)$ and $\tilde{p}_i^{(R)}=\Phi(X_i)$ ($i=1,\ldots,K$), respectively. While for the regular FE, we combine the two-sided $p$-values $p_i=2(1-\Phi(|X_i|))$ for $i=1,\ldots,K$. 

\begin{figure}
\centering
\includegraphics[scale=1]{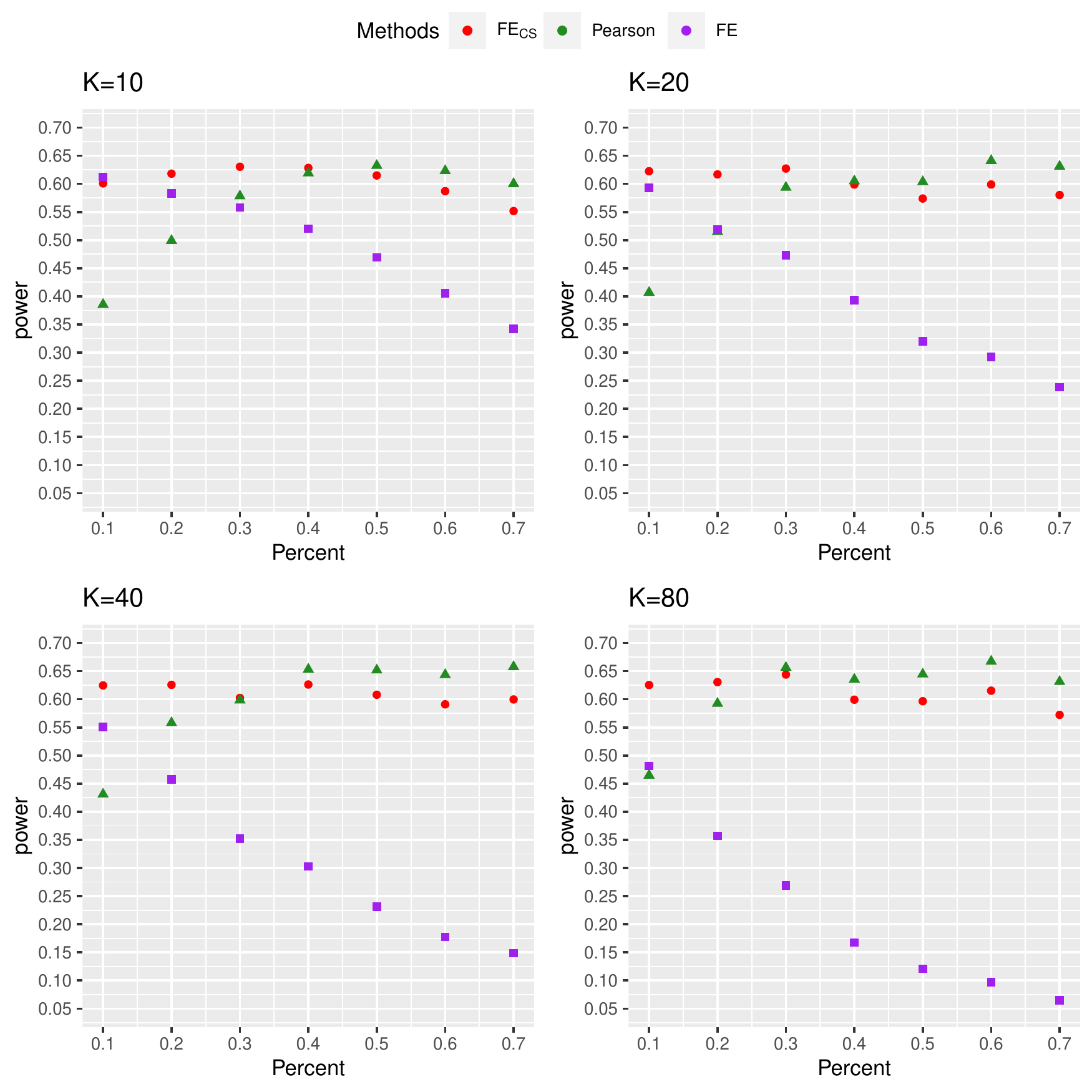}
\caption{Statistical power of FE, FE$_{\text{CS}}$, and Pearson at significance level $\alpha=0.001$ across varying frequencies of signals $\ell/K=0.1,0.2,\ldots,0.7$ and varying numbers of combined $p$-values $K=10,20,40,80$. 
The standard errors are negligible and hence omitted. 
}
\label{power3}
\end{figure}

Figure \ref{power3} shows the empirical power of FE$_{\text{CS}}$, Pearson, and the regular FE at $\alpha=0.001$. For a given frequency of signals $\ell/K$ and a number of combined $p$-values $K$, we choose the smallest $\mu_0$ that allows the best method to have power larger than 0.6. Both FE$_{\text{CS}}$ and Pearson dominate the regular FE, indicating the former two methods perform better for the alternatives with one-sided direction consistent effects (as $\mu_1=\ldots=\mu_s=\mu_0>0$ under the alternatives).
We further note that FE$_{\text{CS}}$ outperforms Pearson when the frequency of signals is low ($\ell/K\leqslant 0.3$), while FE$_{\text{CS}}$ still has a comparative power with Pearson for detecting frequent signals $(\ell/K\geqslant 0.3)$.

\section{Real Application to AGEMAP data}\label{sec:AGEMAP}
In this section, we apply different $p$-value combination methods to analyze the AGEMAP study \citep{zahn2007agemap}. The dataset contains microarray expression of 8,932 genes in sixteen tissues as well as  age and sex variables of 618 mice subjects. We are interested in identifying age-associated marker genes. Following the original paper, we fit the regression model below to detect age-associated genes in each tissue:
$$
Y_{i jk}=\beta_{0 jk}+ \beta_{\text{age},jk}\text {Age}_{ijk}+\beta_{\text{sex}, jk} \text{Sex}_{ijk}+\varepsilon_{i jk} \text{   for $i=1,\ldots,m_{jk}$,}
$$
where $Y_{ijk}$ is the expression level of the $i$-th subject for the $j$-th gene and $k$-th tissue. We consider designs of both two-sided and one-sided tests when combining $p$-values across tissues. In two-sided test design, two-sided $p$-values $(p_{j1}, \cdots,p_{jK})$ for their corresponding $\beta_{\text{age},jk}$ coefficients are combined using Fisher, AFp and FE methods. In this case, the association directions (positive or negative associations) are not considered. By contrast, one-sided test design combines left-tailed $p$-values $(\tilde{p}_{j 1}^{L}, \cdots, \tilde{p}_{j K}^{L})$ or right-tailed $p$-values $(\tilde{p}_{j 1}^{R}, \cdots, \tilde{p}_{j K}^{R})$ respectively using FE$_{\text{CS}}$. Compared to FE, FE$_{\text{CS}}$ is expected to have increased power to detect age-related biomarkers with concordant signals (all positive associated or all negative associated) across tissues while have reduced power for markers with heterogeneous signals (i.e., positive associations in some tissues and negative associations in some others). In this application, both concordant and heterogeneous age-related biomarkers are of interest. FE and FE$_{\text{CS}}$ will serve as complementary tools for different biological objectives.

\begin{figure}
\begin{subfigure}{.5\textwidth}
  \centering
  \caption{}
  \includegraphics[width=1\linewidth]{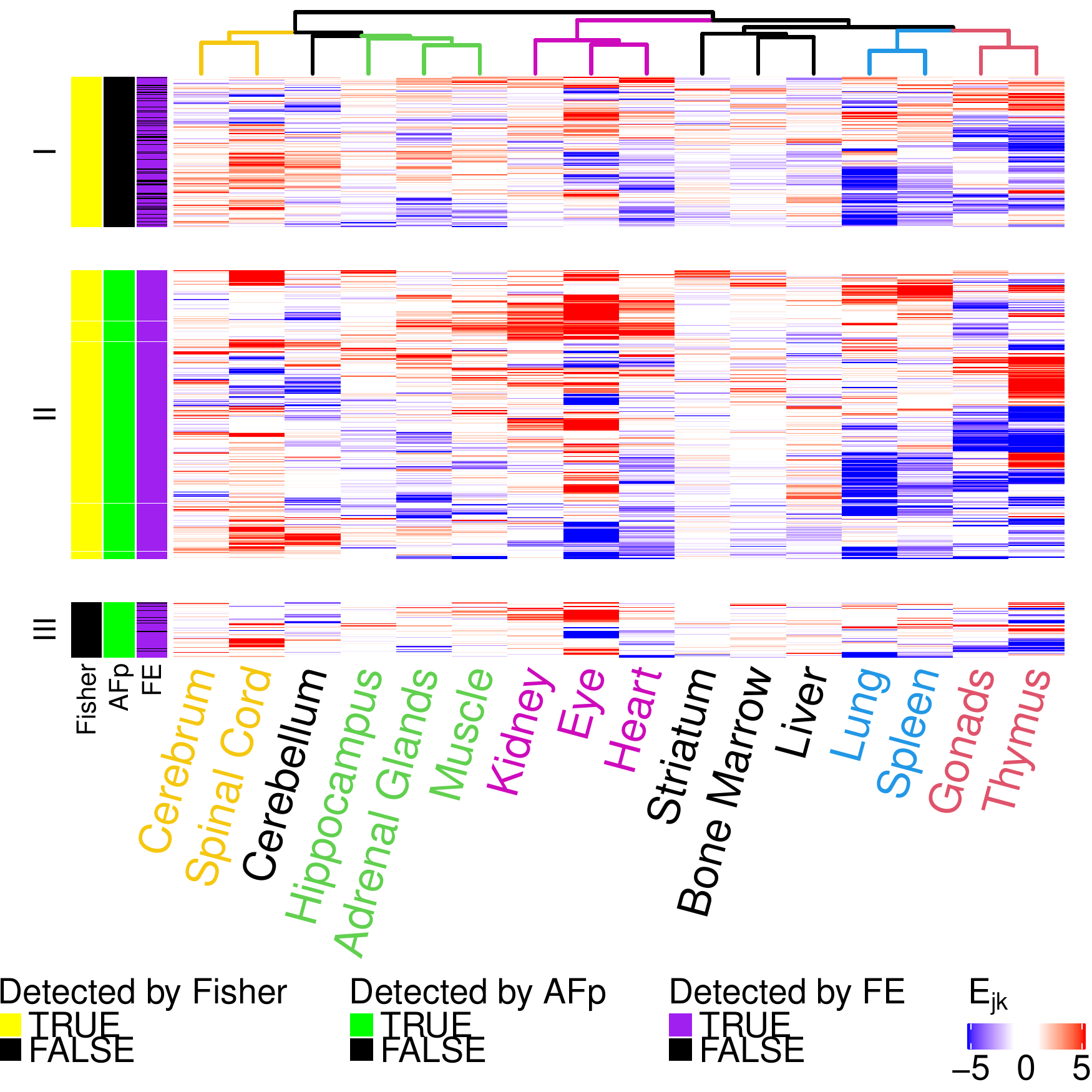}  
  
  \label{fig:sub-first}
\end{subfigure}
\begin{subfigure}{0.5\textwidth}
  \centering
   \caption{}
  \includegraphics[width=1\linewidth]{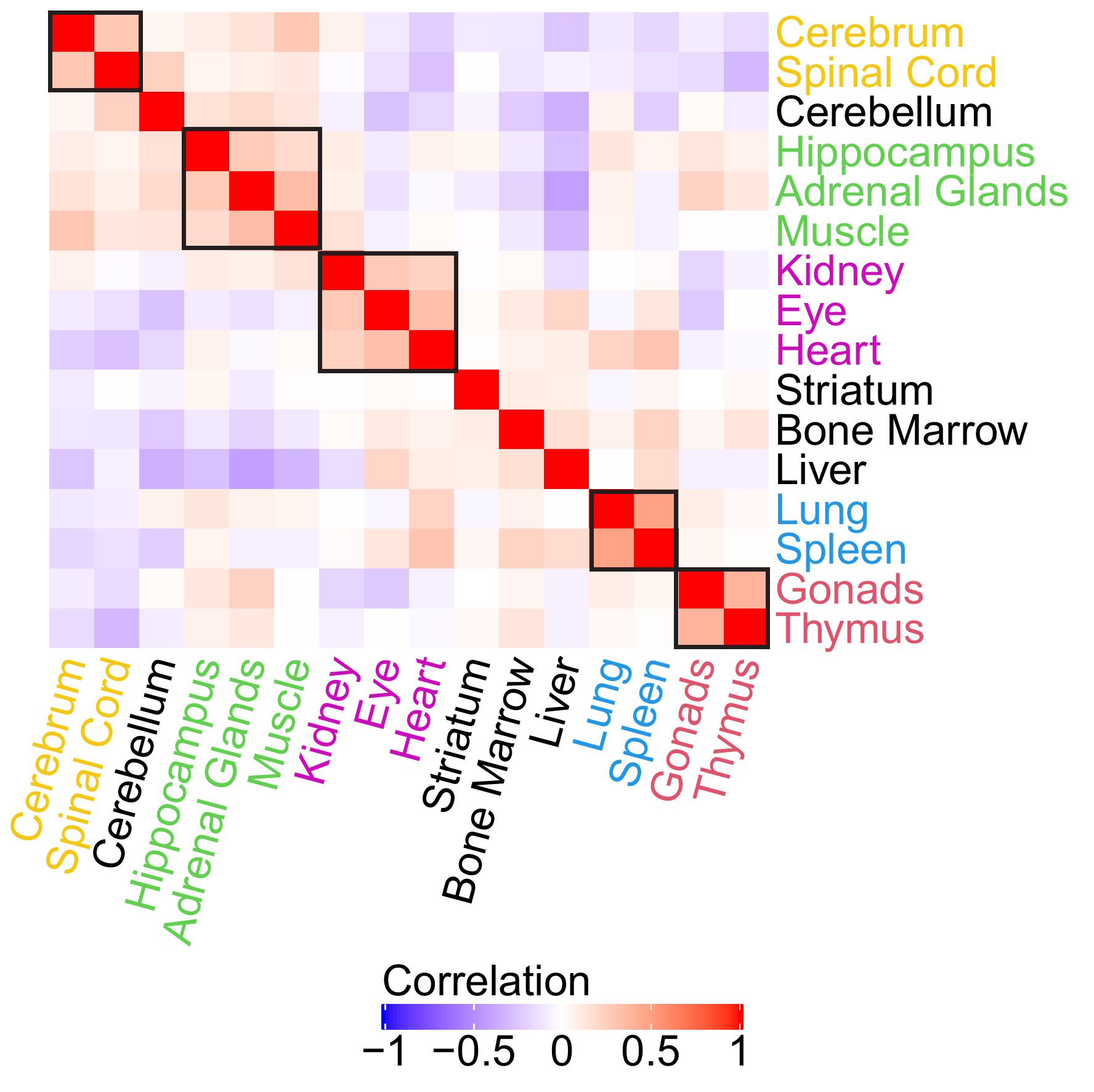}  
  \label{fig:sub-second}
\end{subfigure}
\caption{(a) Heatmaps of age-association measure $E_{jk}$ of significant genes ($q<=0.05$) detected in the two-sided test design. Category I: genes detected by Fisher but not AFp; II: genes detected by both Fisher and AFp; III: genes detected by AFp but not Fisher. (b) Heatmap of pair-wise correlations between tissues based on the detected genes by FE ($q\leqslant 0.05$.) in (a). }
\label{AGEMAPfig:cor and two-sided heatmap}
\end{figure}

Figure \ref{AGEMAPfig:cor and two-sided heatmap}(a) shows Fisher, AFp, and FE $p$-value combination results in the two-sided test design. Under $q$-value$\leqslant 0.05$, Fisher detects 576 genes (yellow color) and AFp detects 452 genes (green color), where Category II (379 genes) represents overlapped detected genes by Fisher and AFp and Categories I (197 genes) and III (73 genes) represent biomarkers uniquely detected by Fisher or by AFp. The heatmap shows age-association measure defined as: $E_{jk}=-\sign(\beta_{\text{age},jk})\log(\min\{\tilde{p}_{jk}^{L},\tilde{p}_{jk}^{R}\})$ for gene $j$ on the rows and tissue $k$ on the columns; i.e., the signed $\log$-transformed (base 10) one-sided $p$-values. Consequently, red color of $E_{jk}$ represents a strong positive association with age while blue means a strong negative association. As expected, FE combines the strengths of Fisher and AFp to detect 598 genes (purple color) that contain all genes in Category II and most genes in Categories I and III. By counting the number of tissues with $p$-values $p_{jk}\leq 0.05$, Supplement Figure S4 shows that category I genes (detected by Fisher but not by AFp) are age-associated in more tissues, while Category III (detected by AFp but not by Fisher) are age-associated in fewer tissues, which is consistent to the theoretical insight and simulation result that Fisher is more powerful for detecting frequent signals and AFp is more powerful for relatively sparse signals.

We next perform hierarchical clustering (using 1-correlation between tissues as dissimilarity measure and complete linkage) for the sixteen tissues based on the $E_{jk}$ values in the 598 age-related genes detected by FE, and the dendrogram is shown in Figure \ref{AGEMAPfig:cor and two-sided heatmap}(a). By cutting the dendrogram, five clear tissue modules of similar age-association patterns are identified: (1) thymus and gonads; (2) spleen and lung;  (3) eye, kidney, and heart (4) hippocampus, adrenal glands, and muscle; (5) cerebrum and spinal cord (also see Figure \ref{AGEMAPfig:cor and two-sided heatmap}(b) for heatmap of pair-wise correlations). For the first module, the thymus has long been regarded as an endocrine organ that is closely related to Gonads and sexual physiology, such as sexual maturity and reproduction. 
\citep{grossman1985interactions,leposavic2018intrinsic}.
The spleen-lung module is consistent with the finding in \cite{zahn2007agemap}, and many reports suggest that spleen and lung share a similar aging pattern (e.g., 
\cite{schumacher2008delayed}). For the third module, literature shows that kidney and eye share structural, developmental, physiological, and pathogenic similarities and pathways. The relationships between age-related eye, kidney, and cardiovascular diseases have been widely reported (e.g., 
\cite{farrah2020eye}).
For the fourth module, extensive literature have reported the relationship between adrenal glands and hippocampal aging (e.g., 
\cite{landfield1978hippocampal}). 
For the last module, few existing studies have investigated the aging process of the spinal cord \citep{knight2017anatomy}. But it is reasonable that the cerebrum and spinal cord might share a similar aging pattern as they both belong to the central nervous system. 
On the other hand, the liver has intriguingly negative correlations of aging effects with muscle, adrenal glands, and several brain regions, such as the hippocampus, cerebellum, and cerebrum (also see Figure \ref{AGEMAPfig:cor and two-sided heatmap}(b)).

\begin{figure}
\centering
\includegraphics[scale=0.75]{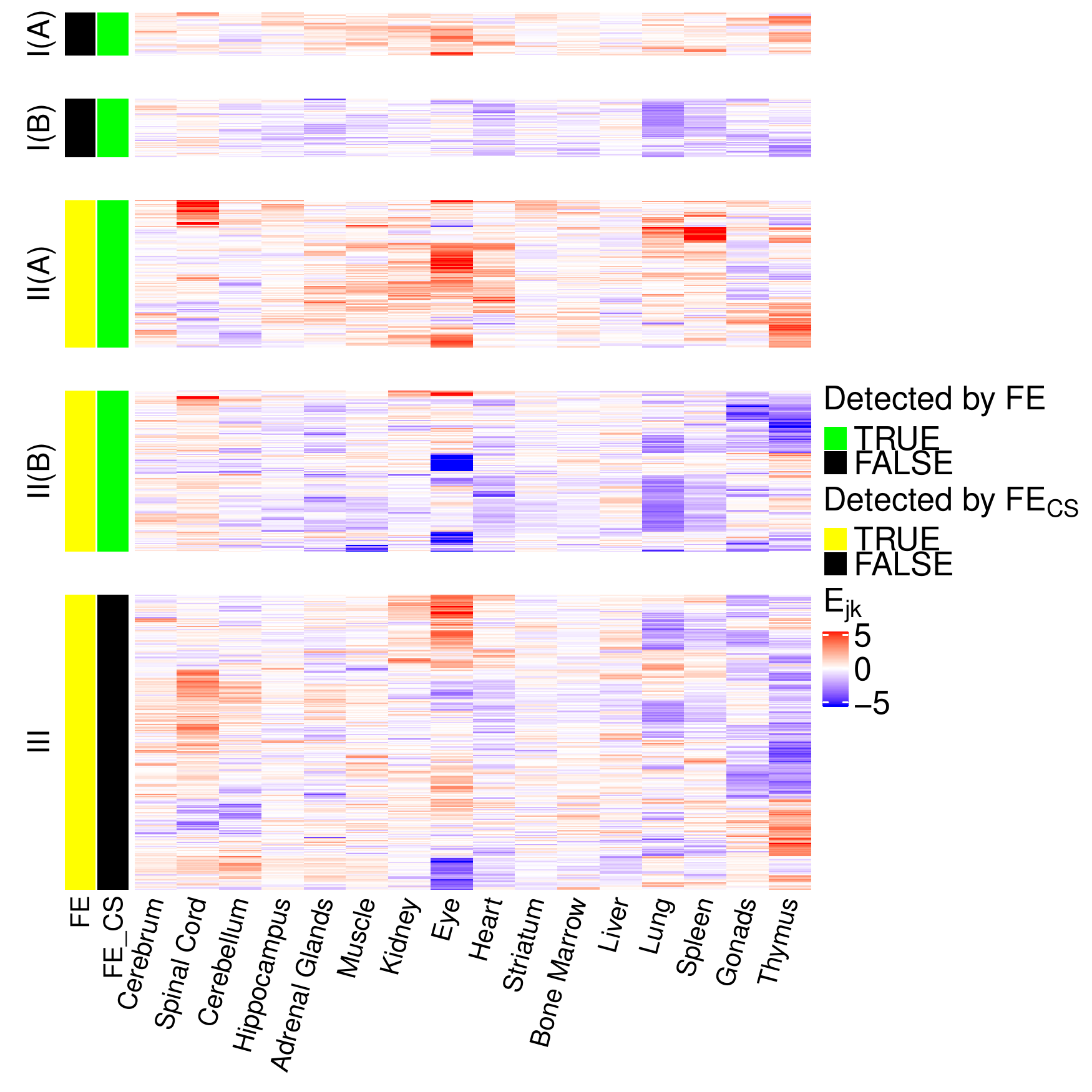}
\caption{Heatmaps of age-association measure $E_{jk}$ of genes detected by FE$_{\text{CS}}$ or by FE ($q<=0.05$). Heatmap I(A) represents up-regulated genes detected only by FE$_{\text{CS}}$ (43 genes); heatmap I(B) represents down-regulated genes detected only by FE$_{\text{CS}}$ (58 genes); heatmap II(A) represents up-regulated genes detected both by FE$_{\text{CS}}$ and FE (146 genes); heatmap II(B) represents down-regulated genes detected both by FE$_{\text{CS}}$ and FE (160 genes);
heatmap III represents genes detected only by FE (292 genes), respectively. 
}
\label{AGEMAPfig:one-sided heatmap}
\end{figure}

Next, we evaluate FE$_{\text{CS}}$ for one-sided test design and compare it with FE. We calculate $S_{\text{sign},j}=\sum_{k=1}^{16}\sign(\beta_{\text{age},jk})\I_{\{\min\{\tilde{p}_{jk}^L,\tilde{p}_{jk}^R\}\leqslant 0.05\}}$ to determine whether the detected concordant aging marker $j$ is positively associated ($S_{\text{sign},j}>0$) or negatively associated ($S_{\text{sign},j}\leqslant 0$) and use it to determine whether a detected marker is dominantly with positive association or negative association. Similar to the previous analysis, Figure \ref{AGEMAPfig:one-sided heatmap} shows age-associated genes detected by FE (598 genes, Categories II(A), II(B) and III) and FE$_{\text{CS}}$ (407 genes, Categories I(A), I(B), II(A) and II(B)), where Categories II(A) and II(B) are overlapped genes detected by FE and FE$_{\text{CS}}$, Category III are only detected by FE and Categories I(A) and I(B) are only detected by FE$_{\text{CS}}$. For genes detected by FE$_{\text{CS}}$, Categories I(A) and II(A) are concordant aging markers with positive association (mostly red) and Categories I(B) and II(B) are negatively associated (mostly blue), which are visually consistent with the heatmap. In contrast, genes in Category III mostly have discordant association directions (partial red and partial blue). {Supplement Figure S5 shows the distributions of $S_{\text{sign},j}$ in the gene categories. } 

At significance level $q\leqslant 0.05$, FE$_{\text{CS}}$ identifies 189 positively associated genes (Categories I(A) and II(A)) and 218 negatively associated genes (Categories I(B) and II(B)). We perform Ingenuity Pathway Analysis (IPA) to these two concordant age-associated gene lists. The result identifies eleven enriched pathways from the 189 positively associated genes and five enriched pathways from the 218 negatively associated genes (enrichment $p\leqslant 0.01$). Table S2 shows details of these enriched pathways with pathway names, enrichment $p$-values, and abundant supporting literature of the pathways related to aging/early development processes (see complete references in Supplement References II). The result shows the advantage of FE$_{\text{CS}}$ to identify age-associated markers concordant across tissues and to deliver interpretable biological insights.

\section{Conclusion and discussion}\label{sec:discussion}
P-value combination is a common and effective information synthetic tool in many scientific applications. In this paper, we focus on a meta-analytic scenario, where the number of combined $p$-values $K$ is relatively small and fixed in the asymptotic setting. Our contribution is three-fold. Firstly, this is the first study to comprehensively evaluate $p$-value combination methods for their asymptotic efficiencies in terms of asymptotic Bahadur optimality (ABO). We investigate classical methods (Fisher and Stouffer) and modified Fisher's methods (AFp, AFz, TFhard, and TFsoft). The result shows that Fisher, AFp, TFhard, and TFsoft are ABO, but Stouffer and AFz are not. We also find an interesting consistency property for the estimation of signal contributing subset in AFp (Theorem \ref{AFpConsistency}). Secondly, we perform an extensive finite-sample power comparison and conclude that Fisher and AFp are two top performers with complementary advantages, where Fisher is more powerful with frequent signals and AFp is more powerful in relatively sparse settings. Thirdly, we propose a Fisher ensemble (FE) method to combine Fisher and AFp. A one-sided test modification, FE$_{\text{CS}}$, is further developed for detecting concordant signals. The advantages of FE and FE$_{CS}$ includes: (A) Both methods have high asymptotic efficiencies (FE is ABO). (B) The truncated Cauchy combination avoids the $-\infty$ score in the traditional Cauchy. (C) We numerically demonstrate their constantly high performance across varying proportions of signals. (D) Both methods have fast-computing procedures. 
Finally, an application to AGEMAP transcriptomic data verifies theoretical conclusions, demonstrates superior performance of FE and FE$_{\text{CS}}$, and discovers intriguing biological findings in age-associated biomarkers and pathways.

Modern data science faces challenges from data heterogeneity, increasingly complex data structure, and the need for effective methods for new scientific hypotheses. The ensemble methods proposed in this paper, FE and FE$_{\text{CS}}$, have solid theoretical and numerical support for their superior performance in a wide range of signal settings. We believe the methods will find impactful applications in many other scientific problems.

\section*{Acknowledgements}
YF and GCT are funded by NIH R21LM012752; CC is funded by Ministry of Science and Technology of ROC 109-2118-M-110-002 and 110-2118-M-110-001. 
The authors thank Yongseok Park for helpful discussion.\vspace*{-8pt}
\bibliographystyle{biom} 
\bibliography{ref}


\end{document}